\documentclass[useAMS,usenatbib]{mn2e}

\usepackage{txfonts}
\usepackage{natbib}
\usepackage[dvips]{graphicx}

\usepackage[ps2pdf=true,pagebackref=true]{hyperref}
\hypersetup{pdftitle=Investigating clustering dark energy with 3d weak cosmic shear}

\def\del#1{{}}

\newcommand{\ltsima}{$\; \buildrel < \over \sim \;$}
\newcommand{\lsim}{\lower.5ex\hbox{\ltsima}}
\newcommand{\gtsima}{$\; \buildrel > \over \sim \;$}
\newcommand{\gsim}{\lower.5ex\hbox{\gtsima}}
\newcommand{\bra}{\langle}
\newcommand{\ket}{\rangle}

\newcommand{\dd}{\mathrm{d}}

\newcommand{\dirac}{\delta_D}

\newcommand{\eg}{e.\,g.}
\newcommand{\ie}{i.\,e.}
\newcommand{\egs}{e.\,g.\/ }
\newcommand{\ies}{i.\,e.\/ }
\newcommand{\cf}{cf.}
\newcommand{\cfs}{cf.\/ }
\renewcommand{\vec}[1]{\ensuremath{\bmath{#1}}}

\title{Investigating clustering dark energy with 3d weak cosmic shear}

\author[Y.~Ayaita, B.~M.~Sch{\"a}fer, M.~Weber]
{Youness Ayaita$^{1}$,
Bj{\"o}rn Malte Sch\"afer$^2$,
and Maik Weber\thanks{e--mail: M.Weber@thphys.uni-heidelberg.de}$^{1}$
\\
$^1$
Institut f{\"ur} Theoretische Physik, Universit{\"a}t Heidelberg,
Philosophenweg 16, 69120 Heidelberg, Germany
\\
$^2$
Astronomisches Recheninstitut,
Zentrum f{\"u}r Astronomie, Universit{\"a}t Heidelberg,
M{\"o}nchhofstra{\ss}e 12, 69120 Heidelberg, Germany}

\begin{document}

\pagerange{\pageref{firstpage}--\pageref{lastpage}}
\pubyear{2011}
\maketitle
\label{firstpage}

\begin{abstract}
	As observational evidence increasingly consolidates the case for a
	cosmological constant $\Lambda$ being the source of the Universe's
	accelerated expansion, the question whether, and if so, how well,
	future experiments could detect deviations from this standard
	scenario is raised with urgency. Assuming a dark energy component
	different from a cosmological constant, the observable effects in
	general include gravitational clustering described by the fluid's
	(rest--frame) speed of sound $c_s$. We employ 3d weak cosmic
	shear, a proposed method to take advantage of the full
	three--dimensional information inherent to the cosmic shear field,
	to explore the capability of future surveys to detect dark energy
	clustering and the signature of an enhanced amplitude of the
	matter power spectrum on large scales. For this purpose, we
	present adequate numerical methods facilitating 3d weak cosmic
	shear calculations. We find that the possible constraints heavily
	depend on the dark energy equation of state $w$. If $w$ is not
	very close to $-1$, constraining the squared sound speed $c_s^2$
	within an order of magnitude seems possible with a combination of
	Euclid and Planck data.
\end{abstract}


%
%

\section{Introduction}
\label{sec:introduction}

Explaining the accelerated expansion of the Universe is one of the key
tasks of cosmology today. If Einstein's general relativity remains
unaltered on cosmological scales, the observed accelerated
expansion---if no local effect---is due to a cosmological constant or
an unknown cosmological fluid with negative pressure, the dark energy.
To this day, all major observations are consistent with a cosmological
constant \citep{Komatsu10, Bartelmann10a}. Its unexpectedly tiny
value---the cosmological constant problem---and the fact that its
energy density is comparable to that of matter just today---the
coincidence problem---\citep[see, \eg,][]{Carroll01} motivate the
search for alternative models of dynamically evolving dark energy.

Due to the lack of observational evidence for inhomogeneities in the
dark energy, most studies have only investigated the consequences of a
perfectly homogeneous dark energy component. Such a fluid is
completely determined by its energy density and its equation of state
$w$. Its direct effect is restricted to the expansion history, which
indirectly causes a scale--independent modification of the growth rate
of matter perturbations. In general, however, a cosmological fluid
can also leave signatures, possibly scale--dependent, by virtue of its
perturbations. This could be, in principle, a means to discriminate
between different dark energy models.

Once we include linear perturbations, another characteristic quantity
enters the scene, the sound speed $c_s$. This quantity defines a sound
horizon such that scales outside and inside this horizon can
undergo different evolutions. In general, both $w$ and $c_s$ are
necessary to describe the observable effects of a fluid. In order to
explore the nature of the dark energy, cosmology has to constrain both
$w$ and $c_s$ \citep{Hu01a, Erickson01, DeDeo03, Hu04}.

The remarkable progress of observational cosmology in determining the
fundamental parameters describing our Universe has not yet led to
significant constraints on the dark energy sound speed $c_s$
\citep{Bean04, dePutter10, Li10}. In this work, we study whether next
generation precision observations of the cosmic microwave background
together with the proposed method 3d weak cosmic shear
\citep{Heavens03} have the potential of providing significant progress
in this respect.

3d weak cosmic shear is a method to gain precision information about
the growth of perturbations \citep{Heavens03, Castro05, Kitching11}.
Contrary to ordinary galaxy surveys, it has---like weak lensing in
general---the advantage of being independent of galaxy bias models.
Only well--understood general relativity is needed from the
theoretical side. This is one reason why weak gravitational lensing,
since its beginnings \citep{VanWaerbeke00}, has advanced along with
the CMB to one of the cosmological probes with the largest potential
\citep{Huterer10, Bartelmann10}. Weak lensing methods have in fact
proved to be powerful tools to constrain dark energy, \ies mainly its
equation of state parameter $w$ \citep{Huterer02, Jain03, Heavens03,
Bernstein04, Takada04, Hannestad06, Heavens06, Amendola08,
Hollenstein09, Kilbinger09, Huterer10}. 

Most weak lensing studies consider the case of tomographic
measurements where the sample of lensed galaxies is split up into
redshift bins on which the standard weak lensing methods are applied
\citep{Hu99a, Hu02}. The advantage of tomography is an enhanced
sensitivity due to reduced averaging along a line of sight compared to
unbinned cosmic shear spectra, but the shape of the dark matter power
spectrum is not measured independently from growth factors and
geometry \citep[recent studies about tomography and the relation to 3d
weak lensing include][]{Kitching11, Schaefer11}.

The 3d version of weak lensing is a complement to standard 2d weak
lensing with the aim of retaining the full three--dimensional
information contained in the cosmological shear field. The starting
point is to not only make use of the angular positions of lensed
galaxies on the sky, but to also include their redshifts as a distance
measure such that each individual galaxy provides a measure of the
tidal shear. 

Let us briefly explain our motivation to look specifically into 3d weak lensing
as opposed to tomographic methods. Weak lensing spectra provide an integral
measure of the dark matter power spectrum, weighted with the lensing efficiency
function. The enhancement of the matter power spectrum due to the clustering of
dark energy is restricted to large scales and would thus influence a weak
lensing convergence spectrum only little. A 3d method, however, provides a
direct measurement of the amplitude of the dark matter spectrum and would be
better suited to distinguish enhanced spectra from unenhanced spectra and
therefore to provide constraints on the properties of dark energy and its
clustering. This would effectively break the degeneracy between the power
spectrum shape and the lensing efficiency, consisting of the growth function and
geometrical factors, such that the signature of dark energy induced clustering
should be easier to observe.

The organisation of this paper is as follows. We first describe
clustering dark energy in general, make contact with prominent dark
energy models, and introduce a parameterisation in
Sec.~\ref{sec:clustering}. We then explain the 3d weak cosmic shear
method in Sec.~\ref{sec:3d}. A brief description of the Fisher matrix
method for forecasting parameter constraints is given in
Sec.~\ref{sec:parameter}. We present adequate and efficient numerical
tools in Sec.~\ref{sec:method}. Our results are shown in
Sec.~\ref{sec:results}, and we conclude in Sec.~\ref{sec:summary}.

%
%

\section{Clustering dark energy}
\label{sec:clustering}

\subsection{The sound speed}
\label{sec:sound}

The dynamics of the background and the evolution of scalar linear
perturbations of a cosmological fluid are fully determined by its
equation of state $w = \bar p / \bar \rho$ and its (squared) sound
speed $c_s^2 = \delta p / \delta \rho$. If we describe dark energy as
a cosmological fluid, coupled to other fluids only by virtue of the
gravitational interaction, the natural parameters are $w$ and $c_s^2$.

The sound speed $c_s^2$ defines a characteristic scale $\lambda
\propto |c_s|$, below which the fluid resists gravitational collapse.
In turn, this means that the effects of gravitational clustering are
only observable if the scale $\lambda$ lies within the Hubble
horizon, $\lambda \lesssim \mathcal H^{-1}$, where $\mathcal H =
a'/a$ is the conformal Hubble parameter and a prime denotes a
derivative with respect to conformal time $\tau$.

In general, the speed of sound is defined by the quotient of the
pressure and density perturbations, $c_s^2 \equiv \delta p / \delta
\rho$. Both, $\delta p$ and $\delta \rho$, however, are
gauge--dependent quantities, whence we shall only consider the
gauge--invariant rest--frame speed of sound defined in a frame where
the velocity perturbation of the fluid vanishes, $\upsilon = 0$.

We can illustrate the role of the sound speed with the help of the
evolution equations of linear perturbations, which are obtained from
the general energy--momentum conservation equations
${T^{\mu\nu}}_{;\nu} = 0$. These equations are valid if there is no
coupling, \ies no energy--momentum exchange, between the fluid and
other components such as matter. As usual, we split into background
quantities
and linear perturbations, $T^{\mu\nu} = {\bar T}^{\mu\nu} + \delta
T^{\mu\nu}$, and we define ${T^0}_0 = -\bar \rho(1 + \delta)$,
${T^0}_i = (\bar \rho + \bar p) \upsilon_i$, ${T^i}_j = (\bar p +
\delta p) \delta^i_j + {\Sigma^i}_j$. We further define a
gauge--invariant density perturbation $\Delta = \delta + 3 (1 + w)
\frac{\mathcal H}{k} (\upsilon - B)$ in Fourier space, where $B$ is a
metric perturbation defined as in \citet{Kodama84}. Choosing the
fluid's rest frame $\upsilon = 0$ and a gauge where $B = 0$, we simply
get $\Delta = \delta$. Describing the evolution of perturbations, for
a single fluid, in terms of the variable $\rho a^3 \delta$, we find
the following second--order differential equation \citep[\cf,
\eg,][]{Kodama84},
\begin{equation}
	(\rho a^3 \delta)'' + \left( 1 + 3 \frac{\dot{\bar p}}{\dot{\bar
	\rho}} \right) \mathcal H (\rho a^3 \delta)'
	+ \left( k^2 c_s^2 - \frac{3}{2} (1+w) \mathcal H^2 \right) (\rho
	a^3 \delta) = 0,
\end{equation}
neglecting anisotropic shear, ${\Sigma^i}_j = 0$. A critical scale
$k_\mathrm{crit} \equiv 1 / \lambda_\mathrm{crit}$ is given by the
vanishing of the source term $\propto \rho a^3 \delta$ driving
gravitational collapse, \ie,
\begin{equation}
	\lambda_\mathrm{crit} = \sqrt{\frac{2}{3}} \frac{1}{\sqrt{1 + w}}
	\frac{|c_s|}{\mathcal H}.
	\label{eq:horizon}
\end{equation}
The perturbation variable $\rho a^3 \delta$ can only grow on
subhorizon scales for $\lambda_\mathrm{crit} \lesssim {\mathcal
H}^{-1}$, which translates into the approximate relation $c_s^2
\lesssim 1 + w$. Especially for an equation of state $w$ close to
$-1$, as preferred by current observations \citep{Komatsu10}, this
only occurs for very small sound speeds $c_s^2 \ll 1$. These
effects are restricted to large scales $\lambda \gtrsim
\lambda_\mathrm{crit}$.

In a complete description of the perturbation evolution, we have to
cope with the multi--component fluid of (at least) matter and dark
energy. Nonetheless, we can still motivate a corresponding heuristic
definition of an effective scale characterising dark energy
clustering, see Sec.~\ref{sec:parameterised}. We will then also show
quantitatively how a clustering dark energy component (with constant
$w$ and $c_s^2$) affects the large--scale matter power spectrum
$P(k)$.

The (rest--frame) sound speed $c_s^2$ considered here may not be
confused with the adiabatic sound speed $c_a^2$, which is only equal
to the quotient $\delta p / \delta \rho$ for adiabatic perturbations,
\ies when the entropy perturbation is zero. In general, it is given by
$c_a^2 = \dot{\bar p} / \dot{\bar \rho}$. The difference between the
two quantities defines a gauge--invariant entropy perturbation $(c_s^2
- c_a^2) \delta/w$. For a fluid with constant equation of state $w$,
the adiabatic sound speed simply reduces to $c_a^2 = w$. For a brief
introduction to dark energy clustering, see \citet{Gordon04}.

\subsection{Relation to common dark energy models}
\label{sec:relation}

\subsubsection{Quintessence}
\label{sec:quintessence}

The most prominent example of dynamical dark energy is standard
quintessence \citep{Wetterich88, Ratra88}, \ies a cosmological scalar
field $\varphi$ with standard kinetic term and a potential
$V(\varphi)$, defining a Lagrangian density $\mathcal L = -\frac{1}{2}
\partial_\mu \varphi \partial^\mu \varphi - V(\varphi)$.  For suitable
choices of the potential $V(\varphi)$, the dynamics of the background
field $\bar \varphi$ shows appealing tracker behaviours providing
robustness against initial conditions.

The perturbation $\delta \varphi$ of the quintessence field usually is
of little importance on subhorizon scales, the reason being that the
quintessence sound speed $c_s^2$ is unity.

This is easily seen by explicitly writing energy density and
pressure perturbations of the scalar field,
\begin{eqnarray}
	\delta \rho_\varphi &=& \dot \varphi \, \delta\dot{\varphi}
	+ V_{,\varphi} \delta \varphi,
	\label{eq:quint_drho}
	\\
	\delta p_\varphi &=& \dot \varphi \, \delta\dot{\varphi}
	- V_{,\varphi} \delta \varphi.
	\label{eq:quint_dp}
\end{eqnarray}
Since the velocity perturbation $\upsilon$ is proportional to the field
perturbation $\delta \varphi$, the rest--frame speed of sound (for
$\upsilon = \delta\varphi = 0$) is $c_s^2 = \delta p_\varphi / \delta
\rho_\varphi = 1$.

We conclude that the detection of a dark energy sound speed $c_s^2 <
1$ would not only challenge the $\Lambda$CDM model but
standard quintessence models as well.

A class of models with very different behaviour, however, is given by
coupled quintessence models \citep{Wetterich95, Amendola00,
Amendola08a}. In these models, there is an energy--momentum exchange
between the dark energy and other components such as dark matter or
neutrinos. Dark energy can then no longer be described as an
independent fluid, and the equations of Sec.~\ref{sec:sound} do not
apply. In fact, subhorizon perturbations of the quintessence field can
grow in these models.  Although not considered in this work, the case
of energy--momentum exchange between dark energy and matter has been
parametrised and studied in the light of weak lensing
\citep{Schaefer08, LaVacca08, CalderaCabral09, deBernardis11}.

\subsubsection{$k$--essence}
\label{sec:kessence}

Looking at Eqs.~(\ref{eq:quint_drho}) and (\ref{eq:quint_dp}), the
reason for $c_s^2 = 1$ in standard quintessence is the identical
dependence of $\delta\rho_\varphi$ and $\delta p_\varphi$ on
$\delta \dot{\varphi}$. Formally, this could easily be changed by
allowing the potential to also depend on $\dot \varphi$, $V =
V(\varphi, \dot \varphi)$. If this dependence can be split into two
summands, we could reinterpret the $\dot \varphi$ dependence as a
modification not of the potential but of the kinetic term.

Non--standard kinetic terms are the starting point for $k$--essence
models of dynamical dark energy \citep{ArmendarizPicon00,
ArmendarizPicon00a}. In these models, the Lagrangian
$\mathcal L$ is a generic function of the standard kinetic term $X =
-\frac{1}{2} \partial_\mu \varphi \partial^\mu \varphi$. It is thus
possible for the sound speed $c_s^2$ to take any value, without
violating causality \citep{Babichev07}.

The energy density and the pressure are given by the
corresponding components of the energy--momentum tensor. They read
$\rho = 2 \mathcal L_{,X} X - \mathcal L$ and $p = \mathcal L$. The
equation of state $w = \bar p/ \bar \rho$ and the rest--frame sound
speed $c_s^2 = \delta p / \delta \rho$ \citep{Erickson01} are then
\begin{eqnarray}
	w &=& \frac{\mathcal L}{2 \mathcal L_{,X} X - \mathcal L},\\
	c_s^2 &=& \frac{\mathcal L_{,X}}{\mathcal L_{,X} + 2 \mathcal
	L_{,XX} X}.
\end{eqnarray}
Of course, both $w$ and $c_s^2$ evolve in time and may take very
different values at different epochs. The question whether the
time evolution of $c_s^2$ could leave characteristic observational
imprints was studied by \citet{Ansari11}. In the framework of a specific
$k$--essence type model, 3d cosmic shear has been used to forecast
possible constraints on the model parameters \citep{Camera10}.

\subsection{Parametrised clustering dark energy}
\label{sec:parameterised}
\label{sec:wcdm}

A frequently employed parameterisation of dark energy, which we shall
adopt here, is the $w$CDM model, sometimes called $X$CDM
\citep{Turner97}. Contrary to a cosmological constant $\Lambda$ with
equation of state $w_\Lambda = -1$, the model allows for an arbitrary
dark energy equation of state $w$ which is taken to be constant in
time. The model is often extended to a linear evolution of $w$ with
respect to the scale factor $a$ \citep[for an attempt to study $w$ as
a free function, \cf][]{Huterer01}. The simplest generalisation for
including possible clustering of dark energy is to further introduce a
rest--frame sound speed $c_s^2$, also constant in time. In this paper,
we completely parametrise the dark energy component by constant
numbers $w$ and $c_s^2$.

Dynamical dark energy such as quintessence and $k$--essence provides
a large class of models that cannot be approximated by a simple
parameterisation such as the $w$CDM model. In fact, the $w$CDM model
(for constant $w$) does not resemble very closely {\it any} of the
prominent dynamical models. Whenever new observational data are
published, it is thus not sufficient to study constraints in the
$w$CDM model alone, but to also study the individual dark energy models.

Nonetheless, the $w$CDM model is, in terms of its parameters, a
somewhat minimal extension of the standard $\Lambda$CDM model,
including the latter as a special case. Hence, it is a useful tool to
forecast how strong the deviations from $\Lambda$CDM must be for
future observations to detect them.

We now turn to the description of linear perturbations in the presence
of a clustering dark energy component parametrised as above. The
linear growth of perturbations is described by a growth function
$g(k,a)$ that links the Newtonian gravitational potential $\Phi_k(a)$
at scale factor $a$ to the one today $\Phi_k^0$ according to
$\Phi_k(a) = g(k,a)\, \Phi_k^0 / a$. We shall now introduce the
parameterisation for $g(k,a)$ that we use for the study of dark energy
perturbations in the $w$CDM model.

Dark energy perturbations contribute to the
gravitational potential just as matter perturbations via the Poisson equation,  
\begin{equation}
	k^2 \Phi = - 4 \pi G a^2 \left( \rho_m \Delta^{(m)} +
	\rho_{DE} \Delta^{(DE)} \right) \equiv - 4 \pi G a^2 \, Q \, \rho_m \Delta^{(m)},
	\label{eq:poisson_Delta}
	\label{eq:poisson_q}
\end{equation}
where we have used the gauge--invariant density perturbations $\Delta^{(m)}$ and
$\Delta^{(DE)}$ and introduced the quantity $Q = Q(k,a)$. It is defined via
\begin{equation}
	Q = 1 + \frac{\rho_{DE} \Delta^{(DE)}}{\rho_{m}\Delta^{(m)}}. 
	\label{eq:qdef}
\end{equation}
An important effect of dark energy perturbations is their influence on
the growth of matter perturbations expressed in terms of a modified
growth index $\gamma$ defined by $\dd \ln \Delta^{(m)} / \dd \ln a =
\Omega_m^\gamma$ \citep{Linder07}. As a function of $Q$,
we may approximate \citep[\cf][]{Sapone09, Sapone10, Linder07}
\begin{equation}
	\gamma \approx \frac{3\left( 1- w - A  \right)}{5
	- 6 w},\ A \equiv \frac{Q-1}{1-\Omega_m}.
	\label{eq:gamma_q}
\end{equation}
We follow \citet{Sapone10}, parametrising $Q(k,a)$ for the $w$CDM
model with sound speed $c_s^2$ as
\begin{equation}
	Q \approx 1+\frac{1-\Omega_m^0}{\Omega_m^0}\frac{(1+w)\,
	a^{-3w}}{1-3w+y^2 },\ y^2 \equiv \frac{2}{3}\frac{k^2 c_s^2
	a}{\Omega_m^0 {\mathcal H}_0^2}.
	\label{eq:q_par}
\end{equation}
Together, these equations provide a convenient way of obtaining the
growth function 
\begin{equation}
	g(k,a) = \frac{Q(k,a)}{Q^0(k)}\, \exp \left(\int_{1}^{a} \frac{\dd a'}{a'}\, 
	\Omega_m(a')^{\gamma(k,a')}\right). 
	\label{eq:growth_param}
\end{equation}
The explicit appearance of $Q$ in this expression is due to our definition of
$g$ describing the growth of the total
gravitational potential rather than of the matter perturbations only. 

\begin{figure}
	\begin{center}
		\includegraphics[width=.45\textwidth]{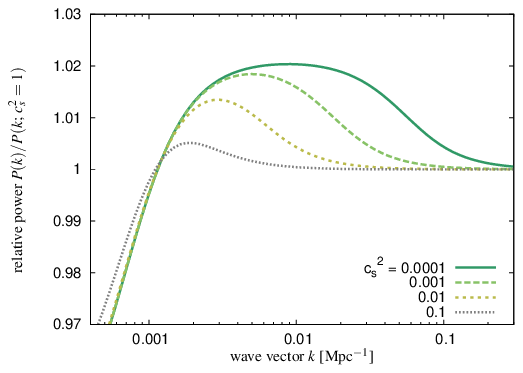}
		\includegraphics[width=.45\textwidth]{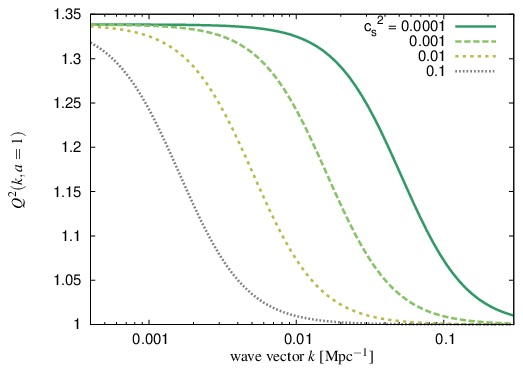}
	\end{center}
	\caption{The upper figure shows the matter power spectrum $P(k)$ for $w =
	-0.8$ and varying
	dark energy speed of sound $c_s^2$ divided by the spectrum for
	$c_s^2=1$. The matter power spectrum is computed from the
	gauge--invariant density contrast $\Delta^{(m)}$. The lower figure shows the
	scale dependence of $Q^2$ for the same values of $w$ and $c_s^2$.}
	\label{fig:relative_spectra}
\end{figure}

Of course, the growth function could easily
be directly obtained by solving the linear perturbation equations
numerically. For illustration, we show the linear matter power
spectrum $P(k)$ for the $w$CDM model for different sound speeds
$c_s^2$ and $w = -0.8$ in Fig.~\ref{fig:relative_spectra}.  Here, we
have used the code {\sc camb} \citep[Code for Anisotropies in the Microwave
Background,][]{Lewis00}, which has built--in facilities to work with
the $w$CDM model with constant $c_s^2$. We have assumed adiabatic initial
conditions.

Perturbations in the dark energy act as an extra source of the
gravitational potential in the Poisson equation~(\ref{eq:poisson_q})
enhancing the growth of matter perturbations on subhorizon scales.
This enhancement, however, is less than a percent effect for sound speeds
$c_s^2 \gtrsim 0.1$ and restricted to large scales. The power
spectrum $P(k)$ has to be known with very high precision in order to
find significant constraints on $c_s^2$. This becomes even more
difficult for $w$ closer to $-1$, \cfs Eq.~(\ref{eq:horizon}). Note
that the plot also shows superhorizon scales, where the results are
gauge--dependent. For our analysis, we will use (subhorizon) scales $k$ between $10^{-3}$ and
$10^{-1}$ Mpc$^{-1}$.  


With the parameterisation of $Q(k,a)$, Eq.~(\ref{eq:q_par}), at hand, we can ask above
which scale $\lambda_\mathrm{eff}$ dark energy clustering could leave
observable traces. Let us make the heuristic assumption that the
effect of a clustering dark energy component would be observable once
roughly $Q(k, a = 1) \gtrsim 1 + \varepsilon$, with $\varepsilon$ for
example at the percent level. This is the case for scales
\begin{equation}
	\lambda \equiv \frac{1}{k} \gtrsim
	\left( \frac{\varepsilon}{1 - \Omega_m^0}
	\right)^{1/2}
	\,\sqrt{\frac{2}{3}} \frac{1}{\sqrt{1 + w}}
	\frac{|c_s|}{ {\mathcal H}_0}.
	\label{eq:eff_scale_analytical}
\end{equation}
For the exemplary values $\varepsilon \approx 1\%$ and $\Omega_m^0 \approx
0.3$, this defines a critical scale
\begin{equation}
	\frac{\lambda_\mathrm{eff}}{ {\mathcal H_0}^{-1}} \approx 
	0.1 \, \frac{|c_s|}{\sqrt{1 + w}}
	\label{eq:eff_scale_numerical}
\end{equation}
with a similar behaviour as the scale given in Eq.~(\ref{eq:horizon}).
A precision experiment might be able to detect dark energy clustering
if the effective scale $\lambda_\mathrm{eff}$ lies within the Hubble
horizon.  In particular, the common choice $w = -1$ refers to
$\lambda_\mathrm{eff} \to \infty$. This is already obvious from the
parameterisation (\ref{eq:q_par}) yielding $Q = 1$ on all scales for
$w = -1$. In this case, the sound speed becomes irrelevant and dark
energy clustering cannot be detected.

%
%

\section{3d weak cosmic shear}
\label{sec:3d}

\subsection{Convergence}
\label{sec:convergence}

In the presence of a gravitational lens, the observed image points
$\vec \theta$ of a galaxy differ from their true positions $\vec
\beta$. In a locally linear approximation, the mapping $\vec \theta
\mapsto \vec \beta$ is described by a matrix
\begin{equation}
	\mathcal A = 
	\left(
	\begin{array}{cc}
		1 - \kappa - \gamma_1 & - \gamma_2 \\
		- \gamma_2 & 1 - \kappa + \gamma_1
	\end{array}
	\right).
\end{equation}
The convergence $\kappa$ determines the magnification of the image,
the shear $(\gamma_1, \gamma_2)$ determines its deformation. Both
convergence and shear are calculated from second derivatives of the
lensing potential $\phi$, \eg,
\begin{equation}
	\kappa = \frac{1}{2}\, \Delta_{\vartheta,\varphi} \, \phi.
	\label{eq:kappa_poisson}
\end{equation}
The lensing potential $\phi$ is a projection of the Newtonian
gravitational potential $\Phi$. In a flat universe,
\begin{equation}
	\phi(\vec \chi) = 2 \int_{0}^{\chi} \dd \chi' \, \frac{\chi -
	\chi'}{\chi\chi'} \,
	\Phi(\vec \chi'),
	\label{eq:lensing_potential}
\end{equation}
where $\vec \chi$, $\vec \chi'$ denote comoving coordinates. The
convergence $\kappa$ thus depends on the gravitational potential along
the line of sight, which, in turn, is given by the density
fluctuations. In this way, gravitational lensing can be used to probe
the density field, without relying on galaxy bias models
\citep{Jain97, Hu01}. For general treatments of weak gravitational
lensing, see \citet{Bartelmann01, Bartelmann10}.

In a region of the sky covered by a weak lensing survey, the
individual convergences $\kappa$ and shears $\gamma_i$ of the galaxies
together allow to study the two--dimensional fields
$\kappa(\vartheta,\varphi)$, $\gamma_i (\vartheta, \varphi)$. If the
galaxies' distances are known (\egs by a photometric redshift measurement),
the fields become three--dimensional,
$\kappa(\chi,\vartheta,\varphi)$, $\gamma_i(\chi, \vartheta,
\varphi)$. 3d weak cosmic shear is a means to study the statistical
properties of these fields \citep{Heavens03, Castro05}.

The importance of the three--dimensional information for weak lensing
precision tests of structure formation has first been studied for
tomography \citep{Hu02}. Also the use of spectroscopic redshifts
instead of a photometric method has been considered \citep{Ishak05}.

The statistics of the convergence field $\kappa$ are hardly directly
observable. But since the statistics of convergence and shear are
equivalent, we may use the convergence $\kappa$ instead of $\gamma_i$
in our theoretical calculations.

The first step in a 3d weak cosmic shear calculation is a combined
Fourier and spherical harmonic transform,
$\chi \to k$, $(\vartheta,\varphi) \to (\ell, m)$,
\begin{equation}
	\kappa_{\ell m}(k) = \sqrt{\frac{2}{\pi}} \int_{}^{} \chi^2 \dd \chi \, \dd \Omega \,
	\kappa(\chi,\vartheta,\varphi) \, j_\ell(k\chi) \, Y_{\ell
	m}^*(\vartheta,\varphi).
	\label{eq:harmonictransform}
\end{equation}
By means of the growth factor $g(k,a)$, we may replace the
gravitational potential in Eq.~(\ref{eq:lensing_potential}) by the
potential of today, $\Phi_{\ell m}(k) = g(k,a)\,\Phi^0_{\ell m}(k)/a$.
Note that in clustering dark energy scenarios, the growth factor is
scale--dependent. In the transformed variables,
Eqs.~(\ref{eq:kappa_poisson}), (\ref{eq:lensing_potential}), and
(\ref{eq:poisson_q}) take the simple forms
\begin{eqnarray}
	\kappa_{\ell m} &=& - \frac{\ell(\ell + 1)}{2} \, \phi_{\ell m},
	\label{eq:kappa_poisson_transf}
	\\
	\phi_{\ell m} &=& \eta_\ell (k,k') \, \Phi^0_{\ell m}(k'),
	\label{eq:lensing_potential_transf}
	\\
	k^2 \Phi_{\ell m} &=& 
	 -4\pi G a^2\, Q\, \rho_m \Delta^{(m)}_{\ell m},
	\label{eq:poisson_transf}
\end{eqnarray}
where we have, following \citet{Heavens03}, introduced the quantity
\begin{equation}
	\eta_{\ell}(k,k') = \frac{4}{\pi} \int_{0}^{\infty} \chi^2 \dd
	\chi \, j_\ell(k\chi) \,
	\int_{0}^{\chi} \dd \chi'\, \frac{\chi - \chi'}{\chi\chi'} \,
	j_\ell(k'\chi') \,
	\frac{g(k',a')}{a'}
	\label{eq:eta}
\end{equation}
and used the summation convention
\begin{equation}
	A(k,k')\,B(k',k'') \equiv \int_{0}^{\infty} k'^2 \dd k' \,
	A(k,k')\, B(k',k'').
	\label{eq:summation_convention}
\end{equation}

The appearance of $Q$ on the right--hand side of the Poisson equation accounts
for the direct contribution, $\propto \rho_{DE} \Delta^{(DE)}$, of dark energy
perturbations to the gravitational potential. The indirect and dynamical effect
of dark energy clustering on the evolution of matter perturbations $\rho_m
\Delta^{(m)}$ is accounted for by the modified growth index $\gamma$ as outlined in
Sec.~\ref{sec:parameterised}. 


\subsection{Estimator}
\label{sec:estimator}

In Sec.~\ref{sec:convergence}, we have seen that the convergence
$\kappa_{\ell m}(k)$ is intimately connected to the density
fluctuation field $\delta_{\ell m}^0(k)$ by virtue of
Eqs.~(\ref{eq:kappa_poisson_transf}) to (\ref{eq:poisson_transf}). In
other words, we can, \eg, use the convergence $\bra \kappa_{\ell m}(k)
\, \kappa_{\ell' m'}(k') \ket$ to probe the matter power spectrum
$P^0(k)$.

\citet{Heavens03} has shown how to construct an appropriate estimator for
a weak lensing survey including photometric redshifts. The two main
ingredients of this estimator are:
\begin{enumerate}
	\item The inclusion of the uncertainty of the redshift
		measurement, for simplicity assumed to be a Gaussian with
		width $\sigma_z$ equal for all galaxies,
		\begin{equation}
			p(\chi'|\chi) \, \dd \chi' = \frac{1}{\sqrt{2\pi} \sigma_z}
			\exp\left[ -\frac{\left(z-z'\right)^2}{2 \sigma_z^2} \right]
			\dd z'.
			\label{eq:prob}
		\end{equation}
		We use a typical figure of $\sigma_z=0.02$ \citep{Heavens03}.
		An extension of the formalism allowing for individual redshift
		errors is possible \citep{Kitching11}.
	\item The survey's galaxy distribution encoded in the number
		density $n(\vec \chi) \equiv n(\chi)$ assumed rotationally
		symmetric. It constitutes a statistical weight
		favoring distances $\chi$ where the density of galaxies is
		higher. We use the forecasted shape
		\begin{equation}
			n(z)\,\dd z \propto z^2\exp\left[-\left(\frac{z}{z_0}\right)^\beta
			\right]\, \dd z
			\label{eq:nofz}
		\end{equation}
		for the Euclid survey. Here, we assume $100$ galaxies per square
		arcminute, $z_0=0.64$, and $\beta=3/2$, yielding a 
		median redshift of $z_\mathrm{med}=0.9$ \citep{Amara07}. For
		convenience, we consider the idealised case  
		that the full sky is covered. For a realistic sky coverage
		$f_{\mathrm{sky}} < 1$, the errors scale approximately by
		$f_{\mathrm{sky}}^{-1/2}$.  
\end{enumerate}
We may then define the estimator $\hat \kappa_{\ell m}$ for the
convergence $\kappa_{\ell m}$ in terms of the actual convergences
$\kappa_g$ of galaxies $g$ as the harmonic transform
\begin{equation}
	\hat \kappa_{\ell m}(k) = \sqrt{\frac{2}{\pi}} \sum_{
	\mathrm{galaxies}\ g}^{}
	\kappa_g \, j_\ell(k \chi_g) \, Y_{\ell m}^* (\vartheta_g,
	\varphi_g).
	\label{eq:trueestimator}
\end{equation}
As explained above, the cosmic shear would be better suited for the analysis of
actual observational data.
The expectation value of $\hat \kappa_{\ell m}$ is
\begin{equation}
	\bar{\kappa}_{\ell m}(k) = Z_\ell (k,k') \, 
	M_\ell (k', k'') \, \kappa_{\ell m}(k''),
	\label{eq:estimator}
\end{equation}
with the summation convention~(\ref{eq:summation_convention}) and the
quantities
\begin{eqnarray}
	Z_\ell(k,k') &=& \frac{2}{\pi} \int_{}^{} \chi'^2 \dd \chi' \,
	\int_{}^{} \dd \chi \, p(\chi'|\chi) \, j_\ell(k'\chi) \,
	j_{\ell}(k\chi'),
	\label{eq:z}
	\\
	M_\ell(k,k') &=& \frac{2}{\pi} \int_{}^{} \chi^2 \dd \chi \,
	j_\ell(k\chi) \, j_\ell(k'\chi) \, n(\chi),
	\label{eq:m}
\end{eqnarray}
taking account for the two main ingredients stated above. Abbreviating
the product $B_\ell(k,k') \equiv Z_\ell (k,k'')\,M_\ell
(k'',k''')\,\eta_\ell( k''', k')$, the covariance of 
$\hat \kappa_{\ell m}$ in terms of the matter power spectrum $P^0(k)$ 
reads
\begin{eqnarray}
	S_\ell(k,k') &\equiv&
	\bra \hat {\kappa}_{\ell m}(k) \hat {\kappa}_{\ell m}(k') \ket \nonumber \\
	&=& A^2 \, B_\ell(k,k'') \frac{[Q^0(k'')]^2\,P^0(k'')}{k''^4} B_\ell(k', k'')
	\label{eq:signal}
\end{eqnarray}
with $A = \frac{\ell(\ell + 1)}{2}\,4\pi G \rho^0_m$.

The full covariance $C_\ell(k,k')$ is obtained by adding the shot
noise $N_\ell(k,k') = \frac{\sigma_e^2}{4} M_\ell(k,k')$ with $\sigma_e^2 =
0.1$
\citep{Heavens03}. This neglects the non--zero correlation between the
ellipticities of neighbouring galaxies due to intrinsic alignments
\citep{Heavens00, 2009IJMPD..18..173S}. This small--scale effect,
however, does not affect our analysis of the large--scale consequences
of a dark energy speed of sound. Further systematic effects have been
studied \citep{March11} but, in general, do not seem to have a strong
impact on parameter estimation \citep{Huterer06, Kitching08,
Takada09}.

%
%

\section{Parameter estimation}
\label{sec:parameter}

We apply a standard Fisher information matrix method to investigate
possible future parameter constraints from upcoming weak lensing
surveys. The Fisher information matrix $F_{\mu\nu}$ is a square matrix
whose indices label (cosmological) parameters $p_\mu, p_\nu$. We
choose the parameters $p_\mu, p_\nu \in \{ \Omega_m, A_s, h, n_s, w,
\log_{10} c_s^2 \}$, assuming flatness: $\Omega_{DE} = 1 - \Omega_m$. The Fisher
matrix determines stringent bounds on how precise a parameter $p_\mu$
can be constrained. If all parameters are estimated from the
experimental data, the individual uncertainty $\Delta p_\mu$ does not
go below the Cram\'{e}r--Rao bound, $\Delta p_\mu \geq
\sqrt{(F^{-1})_{\mu\mu}}$ \citep[for an introduction,
see][]{Tegmark97}. The Cram\'{e}r--Rao bound not only applies to individual
parameters, but it also determines optimal confidence regions for a set of
parameters. For two parameters $p_\mu$
and $p_\nu$, the corresponding coefficients of $F^{-1}$ are a
quadratic form defining an error ellipse.
 
Formally, the Fisher matrix is defined via the likelihood $L$,
\begin{equation}
	F_{\mu\nu} = \left< - \partial_\mu \partial_\nu \ln L \right>.
	\label{eq:fisherdef}
\end{equation}
The likelihood $L \equiv L ({\hat \kappa}_{\ell
m}|\vec p)$ is the probability for an experiment to measure the value
${\hat \kappa}_{\ell m}$ for the estimator given cosmological
parameters $\vec p$. 

The cosmological parameters enter the likelihood
in two ways. First, they predict a power spectrum $P^0(k)$ and a growth function
$g(k, a)$, which, by
Eq.~(\ref{eq:signal}), are decisive quantities for the covariance of
the estimator. Second, they define the background evolution and hence
the distance measures entering the quantities $Z_\ell$, $M_\ell$, and
$\eta_\ell$.

If the likelihood $L$ is a multivariate Gaussian in the data with
covariance matrix $\mathbfss C$, the Fisher matrix is given by
\begin{equation}
	F_{\mu\nu} = \frac{1}{2} \mbox{tr}\left[ {\mathbfss{C}}^{-1}
	(\partial_\mu \mathbfss C) \, {\mathbfss{C}}^{-1} (\partial_\nu
	\mathbfss C) \right],
	\label{eq:fishertrace}
\end{equation}
sensitive to the derivatives of the covariance $\mathbfss C$ with
respect to the cosmological parameters.

For our estimator $\hat \kappa_{\ell m}(k)$, the covariance matrix
carries the indices $(\ell, m, k, k')$. Since different modes $\ell$
and $m$ are uncorrelated, the covariance matrix $\mathbfss C$ splits
into blocks. Further, the covariance $C_\ell(k,k')$ from
Sec.~\ref{sec:estimator} is assumed to be independent of $m$ whereby
all $2\ell + 1$ blocks for a given $\ell$ are identical. In terms of
the covariance $C_\ell(k,k')$, we may reformulate
Eq.~(\ref{eq:fishertrace}) to
\begin{equation}
	F_{\mu\nu} = \sum_{\ell = \ell_\mathrm{min}}^{\ell_\mathrm{max}} 
	\frac{2\ell + 1}{2} \mathrm{tr} \left(
	\mathbfss C_\ell^{-1}\, \partial_\mu \mathbfss C_\ell \,
	\mathbfss C_\ell^{-1}\, \partial_\nu \mathbfss C_\ell \,
	\right).
	\label{eq:fishercovariance}
\end{equation}

It should be kept in mind that the Cram\'{e}r--Rao bounds are realistic
estimates 
of the actual constraints only if the likelihood $L$ as a function of the
parameters $\vec p$ is a Gaussian. This is often violated in the case of 
parameters that are difficult to measure and
therefore weakly constrained, such as the sound speed
parameter $c_s^2$. The broad likelihood extends to regions where the dependence of the matter power
spectrum $P(k)$ on $c_s^2$, \cfs Fig.~\ref{fig:relative_spectra}, cannot be
approximated linearly \citep{Ballesteros10}. This also affects the weak lensing convergence
spectrum considered in this work. Figure~\ref{fig:relative_spectra} suggests
that the logarithm $\log_{10}c_s^2$ is a more natural parameter to describe the
reaction of the model to variations in the dark energy speed of sound. We thus
choose $\log_{10}c_s^2$ as a model parameter in our analysis but emphasize that
the Cram\'{e}r--Rao bounds we calculate are only rough estimates of the actual future constraints. 

A very practical feature of the Fisher matrix is its additivity. Given  
Fisher matrices $F_{\mu\nu}^{(A)}$, $F_{\mu\nu}^{(B)}$ for two
independent experiments $A$ and $B$, the joint Fisher matrix providing
the combined parameter constraints is simply
$F_{\mu\nu}^{(A+B)} =
F_{\mu\nu}^{(A)} + F_{\mu\nu}^{(B)}$. This follows directly from the
multiplication of the corresponding likelihoods and the definition of
the Fisher matrix, Eq.~(\ref{eq:fisherdef}). In our case, we can use
this formalism to include prior information from other experiments
than weak gravitational lensing.

As prior information, we use a Fisher matrix
$F^{(\mathrm{CMB})}_{\mu\nu}$ for the cosmic microwave background
(CMB) based on forecasts for the Planck satellite. We include
temperature ($TT$), polarisation ($EE$), and the cross--correlation
spectrum ($TE$). We calculate the Fisher matrix $F^{(\mathrm{CMB})}$
following \citet{Perotto06}. The predicted noisy spectra ${\tilde
C}_\ell^{TT}$ (temperature only), ${\tilde C}_\ell^{EE}$ ($E$--mode
polarisation), and ${\tilde C}_\ell^{TE}$ (cross--correlation) are
encoded in a $3\times 3$ matrix,
\begin{eqnarray}
	{\mathbfss A}_\ell &=& \frac{2}{(2\ell + 1)\, f_\mathrm{sky}} \times \nonumber 
	\\
	&& \times
	\left(
	\begin{array}{ccc}
		\left( {\tilde C}_\ell^{TT} \right)^2 & 
		\left( {\tilde C}_\ell^{TE} \right)^2 & 
		{\tilde C}_\ell^{TE} \, {\tilde C}_\ell^{TT}
		\\
		\left( {\tilde C}_\ell^{TE} \right)^2 & 
		\left( {\tilde C}_\ell^{EE} \right)^2 & 
		{\tilde C}_\ell^{TE} \, {\tilde C}_\ell^{EE}
		\\
		{\tilde C}_\ell^{TE} \, {\tilde C}_\ell^{TT} &
		{\tilde C}_\ell^{TE} \, {\tilde C}_\ell^{EE} &
		\frac{1}{2} \left[ \left( {\tilde C}_\ell^{TE} \right)^2 +
		{\tilde C}_\ell^{TT} \, {\tilde C}_\ell^{EE} \right]
	\end{array}
	\right)
\end{eqnarray}
with a fraction $f_\mathrm{sky}$ of the CMB covered. From this,
we evaluate the Fisher matrix,
\begin{equation}
	F_{\mu\nu}^{(\mathrm{CMB})}
	= \sum_{\ell = 2}^{\ell_\mathrm{max}} \sum_{PP',QQ'}^{}
	\partial_\mu {\tilde C}_\ell^{PP'}\, {\left({\mathbfss A}_\ell^{-1}\right)}_{PP'QQ'}\,
	\partial_\nu {\tilde C}_\ell^{QQ'}
\end{equation}
with the indices $PP', QQ' \in \{ TT, EE, TE\}$.

Our forecast bases on expected properties of the Planck satellite
\citep{Hollenstein09, Knox95}. We adopt the expected instrument
properties as listed in Table~1 of \citet{Hollenstein09}, namely a sky
coverage $f_\mathrm{sky} = 0.65$, a beam width $\theta_\mathrm{FWHM} =
7$~arcmin, temperature noise $\Delta_T = 28$~$\mu$K~arcmin, and
polarisation noise $\Delta_E = 57$~$\mu$K~arcmin. For the numerical
calculation of the theoretically predicted multipoles, we employ CAMB.

%
%

\section{Method}
\label{sec:method}

In principle, we have already collected the ingredients for our 3d
weak lensing calculations, namely the covariance of the estimator,
\cfs Eq.~(\ref{eq:signal}), and the Fisher information matrix,
Eq.~(\ref{eq:fishercovariance}). Due to the presence of multiple
nested integrals, the actual calculation is involved and motivates the
choice of adequate numerical approaches and techniques. We present our
strategies in this section.

\subsection{The quantities $Z_\ell$, $M_\ell$, and $\eta_\ell$}
\label{sec:quantities}

The expectation value ${\bar \kappa}_{\ell m}$ of the 3d convergence
estimator, Eq.~(\ref{eq:estimator}), mainly is the application of $Z_\ell$
(\ref{eq:z}), $M_\ell$ (\ref{eq:m}), and $\eta_\ell$ (\ref{eq:eta}) on
today's gravitational potential $\Phi^0_{\ell m}(k)$,
\begin{equation}
	{\bar \kappa}_{\ell m} \propto Z_\ell(k,k')\,M_\ell(k',k'')\,
	\eta_\ell(k'',k''')\,\Phi^0_{\ell m}(k'''),
\end{equation}
where each multiplication corresponds to a $k$ integration according
to the convention, Eq.~(\ref{eq:summation_convention}). We have introduced
the shorthand $B_\ell(k,k''') \equiv Z_\ell(k,k') M_\ell(k',k'')
\eta_\ell(k'',k''')$ for the product.

\begin{figure}
	\begin{center}		
		\includegraphics[width=.5\textwidth]{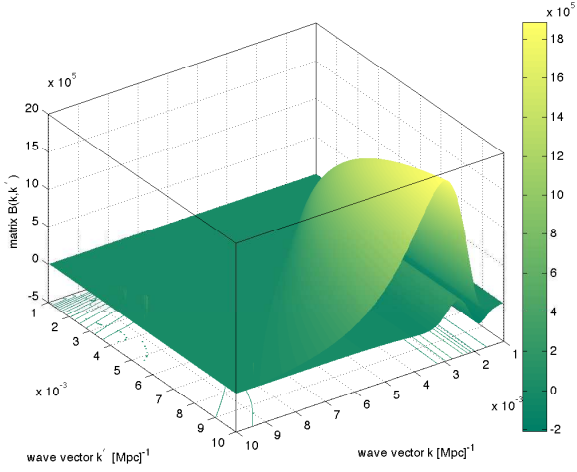}
	\end{center}
	\caption{The matrix $B_\ell(k,k')=
	Z_\ell(k,k'')\,M_\ell(k'',k''')\,\eta_\ell(k''',k')$ for $\ell=10$
	(upper surface) with the difference between the full integration
	and the approximation given by Eq.~(\ref{eq:bfast}) (lower surface
	and contours), which shows a small oscillatory feature close to
	the steep edge of $B_\ell(k,k')$ amounting to less than 10\% of
	the amplitude.}
	\label{fig:b}
\end{figure}

Before we explain an elaborate way to calculate $B_\ell$ with high
precision, we first turn to a simplified approximate approach.
Recalling that the sequence of functions
\begin{equation}
	f_\ell(x) \equiv \sqrt{\frac{2}{\pi}} \,y
	\, \sqrt{\xi}\, j_\ell(\xi), \quad \xi \equiv y (x + 1), \quad y
	\equiv \ell + \frac{1}{2}
\end{equation}
approaches the Dirac delta function $\dirac$ for $\ell \to \infty$, we
may, for sufficiently large $\ell$, use the approximation
\begin{equation}
	j_\ell(k\chi) \approx \sqrt{\frac{\pi}{2}} \, \frac{1}{k \sqrt{y}} \,
	\dirac \left( \chi - \frac{y}{k} \right).
	\label{eq:besselapprox}
\end{equation}
In this approximation, the quantities $Z_\ell$, $M_\ell$, and
$\eta_\ell$ take simple forms, namely
\begin{eqnarray}
	Z_\ell(k,k') &\approx& \frac{y}{k^3 k'} p\left(\left.
	\frac{y}{k} \right| \frac{y}{k'} \right),\\
	M_\ell(k,k') &\approx& \frac{1}{k^2} n\left( \frac{y}{k} \right)
	\dirac (k - k'), \\
	\eta_\ell(k,k') &\approx& 2 \frac{k'-k}{k^3 k'}
	\frac{g(k', a(y/k'))}{a(y/k')} \mbox{ for } k \leq k',\ 0
	\mbox{ else.}
\end{eqnarray}
Calculating the final product $B_\ell$ now does no longer pose
difficulties,
\begin{equation}
	B_\ell(k,k') \approx \frac{y}{k^3 k'} \frac{2
	g(k',a(y/k'))}{a(y/k')}\,
	\int_{\frac{y}{k'}}^{\infty} \dd \chi \, p\left(\left.
	\frac{y}{k} \right| \chi 
	\right) \, n(\chi) \, \frac{\chi - \frac{y}{k'}}{\chi
	\frac{y}{k'}}.
	\label{eq:bfast}
\end{equation}
We compare this approximate result with the full expression in
Fig.~\ref{fig:b}.

Although useful for a first impression, these approximate results do
not allow for a precision calculation of the covariance
$C_\ell(k,k')$. We thus develop a more sophisticated strategy.

\subsection{Covariance}
\label{sec:covariance}
\label{sec:calculation}

The signal and noise parts of the covariance matrix are given in
Sec.~\ref{sec:estimator}. While the noise part $\propto M_\ell$ is
uncomplicated, the direct evaluation of the signal $S_\ell$
(\ref{eq:signal}) would, in a first step, require the calculation of
$Z_\ell$, $M_\ell$, and $\eta_\ell$, which contain highly oscillating
integrands, \cfs Eqs.~(\ref{eq:z}), (\ref{eq:m}), (\ref{eq:eta}). In a
second step, the product $B_\ell(k,k') \equiv Z_\ell (k,k'')\,M_\ell
(k'',k''')\,\eta_\ell( k''', k')$ has to be calculated. Taken
together, these are seven nested integrals. Calculating the signal
covariance $S_\ell$ then requires two further integrations.

Fortunately, the orthogonality relation for spherical Bessel
functions,
\begin{equation}
	\int_{0}^{\infty} k^2 \dd k \, j_\ell(k\chi)\,j_\ell(k\chi') =
	\frac{\pi}{2\chi^2}\dirac (\chi-\chi'),
	\label{eq:bessel_orth}
\end{equation}
can be used to solve several $k$ integrals analytically. The
remaining expression for $B_\ell$ reads
\begin{equation}
	B_\ell(k,k') = \frac{4}{\pi}\int \chi'^2 \dd \chi'\, j_\ell(k\chi')
	\int \dd \chi\, p(\chi'|\chi) \, n(\chi)\, f_{\ell}(k',\chi),
	\label{eq:b}
\end{equation}
where
\begin{equation}
	f_{\ell}(k,\chi) \equiv \int_{0}^{\chi}\dd \chi'\,
	j_\ell(k\chi')\,\frac{\chi-\chi'}{\chi\chi'}\frac{g(k,a')}{a'}.  
	\label{eq:flk}
\end{equation}
The number of nested integrals in the calculation of $B_\ell$ is
reduced to three.

We will show that an efficient evaluation of the inner integral in
Eq.~(\ref{eq:b}) is possible using a Fast Fourier Transform (FFT). We
therefore have to sample $f_\ell$ at discrete coordinates
$\{\chi_j\}$. For each $\chi_j$, we need not calculate the full
integral (\ref{eq:flk}) but only an integral from $\chi_{j-1}$ to
$\chi_j$. This is possible once we write the integral in a way that
makes the integrand independent of the integral bound,
\begin{equation}
	f_\ell(k,\chi) = \int_{0}^{\chi} \frac{\dd \chi'}{\chi'}\,
	j_\ell(k \chi') \frac{g(k,a')}{a'} -
	\frac{1}{\chi}\,\int_{0}^{\chi} \dd \chi' \, j_\ell(k\chi')
	\frac{g(k,a')}{a'}.
\end{equation}
In redshift space, the conditional probability $p(z'|z)$ is a
Gaussian, \cfs Eq.~(\ref{eq:prob}). Inserting this property, we
reformulate the inner integral in Eq.~(\ref{eq:b}) as a convolution,
\begin{displaymath}
	\int_0^\infty \dd \chi\, p(\chi'|\chi) \, n(\chi)\, f_{\ell}(k',\chi)=
\end{displaymath}
\begin{equation}
\quad\quad\quad
	\frac{\dd z'}{\dd \chi'} 
	\int_{0}^{\infty} \dd z \, p(z' - z) 
	\times \left( \frac{\dd \chi}{\dd z} \, n(\chi(z))\, f_\ell(k', z) \right).
\end{equation}
For convolution integrals, fast solving methods exist. This is due to
the convolution theorem stating that the Fourier coefficients of the
individual functions can be multiplied to give the Fourier
coefficients of the convolution. The Fourier transform of the Gaussian
is again a Gaussian and thus analytically known. In the last factor,
we use the sampled values of $f_\ell$ to perform a Fast Fourier
Transform.

\subsection{Fisher matrix}
\label{sec:fisher}

Once the covariance $C_\ell(k,k')$ is known, the Fisher matrix
$F_{\mu\nu}$ can, in principle, be calculated according to
Eq.~(\ref{eq:fishercovariance}). In terms of linear algebra, the
Fisher matrix is given by a trace, which is a basis--independent
operation. This opens the possibility of calculating the covariance in
another basis, allowing for a more efficient numerical calculation.

Working with the tools of linear algebra, we find it more transparent
to abandon the summation convention (\ref{eq:summation_convention})
for a moment and to work with standard notation instead. All earlier
expressions can easily be reproduced if quantities of the type
$A_\ell(k,k')$ are replaced by ordinary matrices
\begin{equation}
	{\mathbfss A}^\ell_{kk'} \equiv \sqrt{k^2 \Delta k} \
	A_\ell(k,k') \, \sqrt{k'^2 \Delta k}
\end{equation}
with a discrete step size $\Delta k$. The additional factors
automatically reproduce the summation convention once a matrix
multiplication is performed, $\sum_{k}^{} k^2 \Delta k \to
\int_{}^{} k^2 \dd k$.

Let us search for an orthogonal transformation ${\mathbfss T}^\ell$ of
the covariance matrix ${\mathbfss C}^\ell_{kk'} = \sqrt{k^2 \Delta k}
\ C_\ell(k,k') \, \sqrt{k'^2 \Delta k}$,
\begin{equation}
	{\tilde{\mathbfss C}}^\ell = \left( {\mathbfss T}^{\ell} \right)^t
	\, {\mathbfss C}^{\ell} \, {\mathbfss T}^\ell.
\end{equation}
A good choice would, when applied on ${\mathbfss C}^\ell$, produce the
orthogonality relation for spherical Bessel functions. Such a choice
is given by
\begin{equation}
	{\mathbfss T}^\ell_{k\rho} = \sqrt{\frac{2}{\pi}} \,
	\sqrt{k^2 \Delta k} \ j_\ell(k\rho) \ \sqrt{\rho^2 \Delta \rho}.
\end{equation}
The noise part ${\mathbfss N}^\ell \propto {\mathbfss M}^\ell$ becomes
particularly simple,
\begin{equation}
	{\tilde{\mathbfss M}}^\ell_{\rho\rho'} =
	\sum_{k,k'}^{} {\mathbfss T}^{\ell}_{k \rho} \,
	{\mathbfss M}^\ell_{kk'} \, {\mathbfss T}^\ell_{k'\rho'}
	= n(\rho) \, \delta_{\rho\rho'}.
\end{equation}
For the transformed signal part, ${\tilde{\mathbfss S}}^\ell = \left(
{\mathbfss T}^{\ell} \right)^t \, {\mathbfss S}^{\ell} \, {\mathbfss
T}^\ell$, the product ${\mathbfss B}^\ell$ is transformed from the
left--hand side only,
\begin{equation}
	{\tilde{\mathbfss B}}^\ell_{\rho k'} = \sum_{k}^{}
	{\mathbfss T}^{\ell}_{k \rho} \,
	{\mathbfss B}^{\ell}_{kk'}.
\end{equation}
In fact, this transformation further simplifies ${\mathbfss B}^\ell$
by virtue of the orthogonality relation,
\begin{eqnarray}
	{\tilde{\mathbfss B}}^\ell_{\rho k'}
	&=& 2 \sqrt{\frac{2}{\pi}} \sqrt{\rho^2 \Delta \rho} \,
	\int_{0}^{\infty} \dd \chi \, p(\rho|\chi)\, n(\chi) \times
	\nonumber \\ && \times
	\int_{0}^{\chi} \dd \chi' \, j_\ell(k'\chi')\,
	\frac{\chi - \chi'}{\chi \chi'} \frac{g(k',a')}{a'}
	\, \sqrt{k'^2 \Delta k}.
\end{eqnarray}

Applying the matrix ${\mathbfss T}^\ell_{k\rho}$ introduced above on a
quantity ${\mathbfss A}^\ell_{kk'}$ can be understood as undoing the
transformation $\chi \to k$ in the harmonic transform, \cfs
Eq.~(\ref{eq:harmonictransform}). This means that the Fourier mode $k$
is replaced by a comoving distance, now labelled by $\rho$. Hence, the
application of ${\mathbfss T}^\ell_{k\rho}$ avoids unnecessary
integrations originating from the harmonic transform.

Finally, we have all the necessary tools for an efficient
calculation of the Fisher matrix at our disposal.

%
%

\section{Results}
\label{sec:results}

The Fisher matrix formalism, \cfs Sec.~\ref{sec:parameter}, and our
numerical methods, explained in Sec.~\ref{sec:method}, enable us to
estimate which constraints on the dark energy sound speed will be
possible with the weak lensing data of Euclid. The constraints depend,
however, on the assumed fiducial parameters since the Fisher matrix is
defined by derivatives at these points (\ref{eq:fishertrace}).
Unfortunately, the dependence of sound speed constraints on the
fiducial values of both the sound speed $c_s^2$ itself and the
equation of state $w$ is very strong. This is illustrated by the scale
$\lambda_\mathrm{eff}$ introduced in Sec.~\ref{sec:parameterised},
Eq.~(\ref{eq:eff_scale_numerical}), below which dark energy clustering
is not expected to be observable.  This scale is a function of both
$c_s^2$ and $w$, it exceeds the Hubble horizon for $c_s^2 \gg 1 + w$. In
particular, the most natural fiducial value for $w$ mimicking the
standard $\Lambda$CDM model, \ie, $w \to -1$, is a singular choice,
$\lambda_\mathrm{eff} \to \infty$. The question of how well the sound
speed $c_s^2$ can be constrained crucially depends on how close the
equation of state $w$ is to the value $-1$.

In order to explore this behaviour quantitatively, we apply the Fisher
matrix formalism to estimate the uncertainties of the dark energy sound speed
and equation of state as functions of the fiducial values $c_s^2$ and $w$. 
In Sec.~\ref{sec:parameter}, we argued that a natural 
parameter to constrain is the order of magnitude $\log_{10}c_s^2$ rather than
$c_s^2$ itself. The relative
error on the sound speed approximately is $\Delta c_s^2/c_s^2 \approx
\ln (10)\, \Delta \log_{10}c_s^2$. This becomes imprecise for large
uncertainties. In the case of the equation of state, we estimate $\Delta w/|w|$. 
For simplicity, we assume all other
cosmological parameters to be exactly known, fixed to the WMAP7
recommended $\Lambda$CDM parameters \citep{Komatsu10}. 

\begin{figure}
	\begin{center}		
		\includegraphics[width=.5\textwidth]{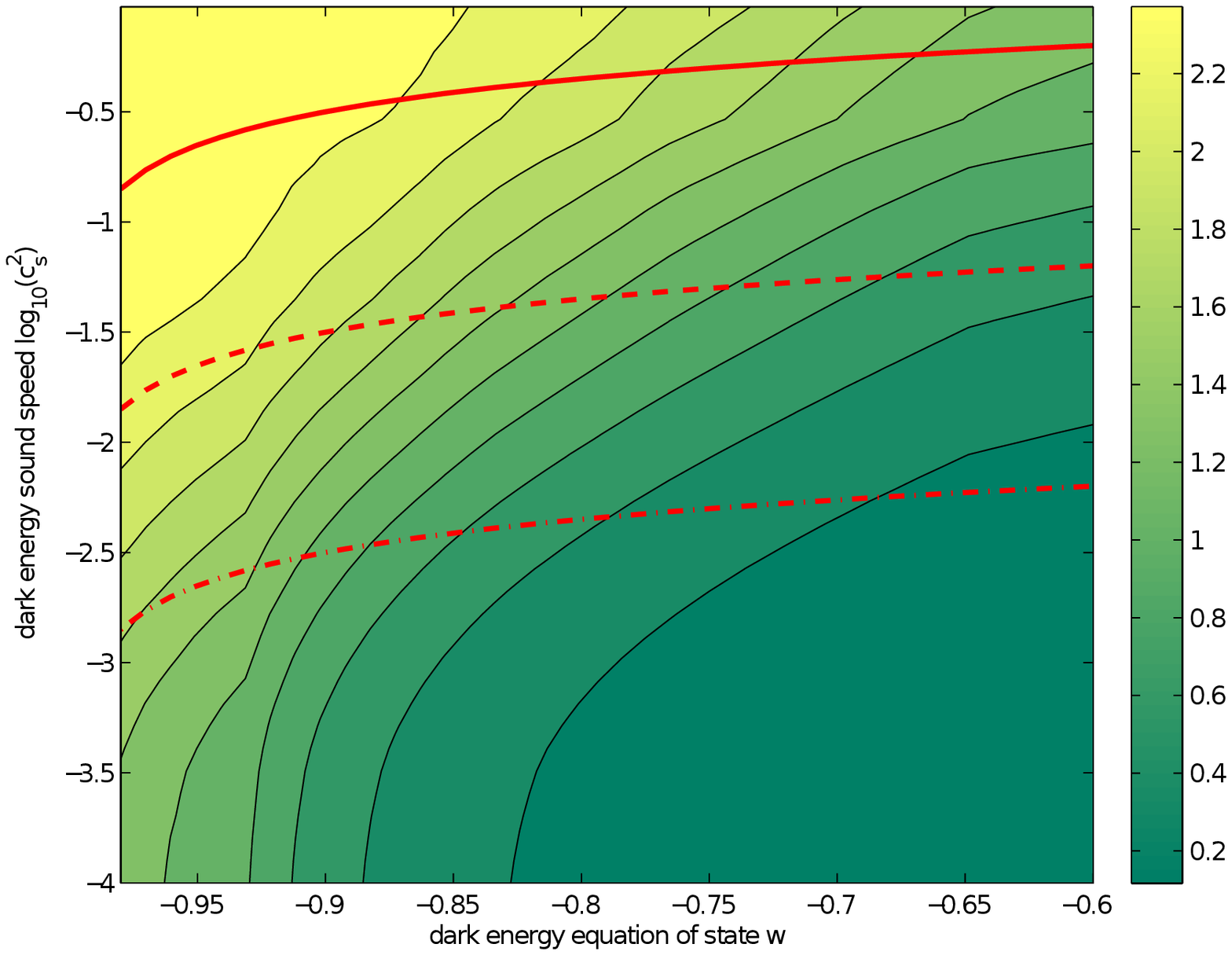}
		\includegraphics[width=.5\textwidth]{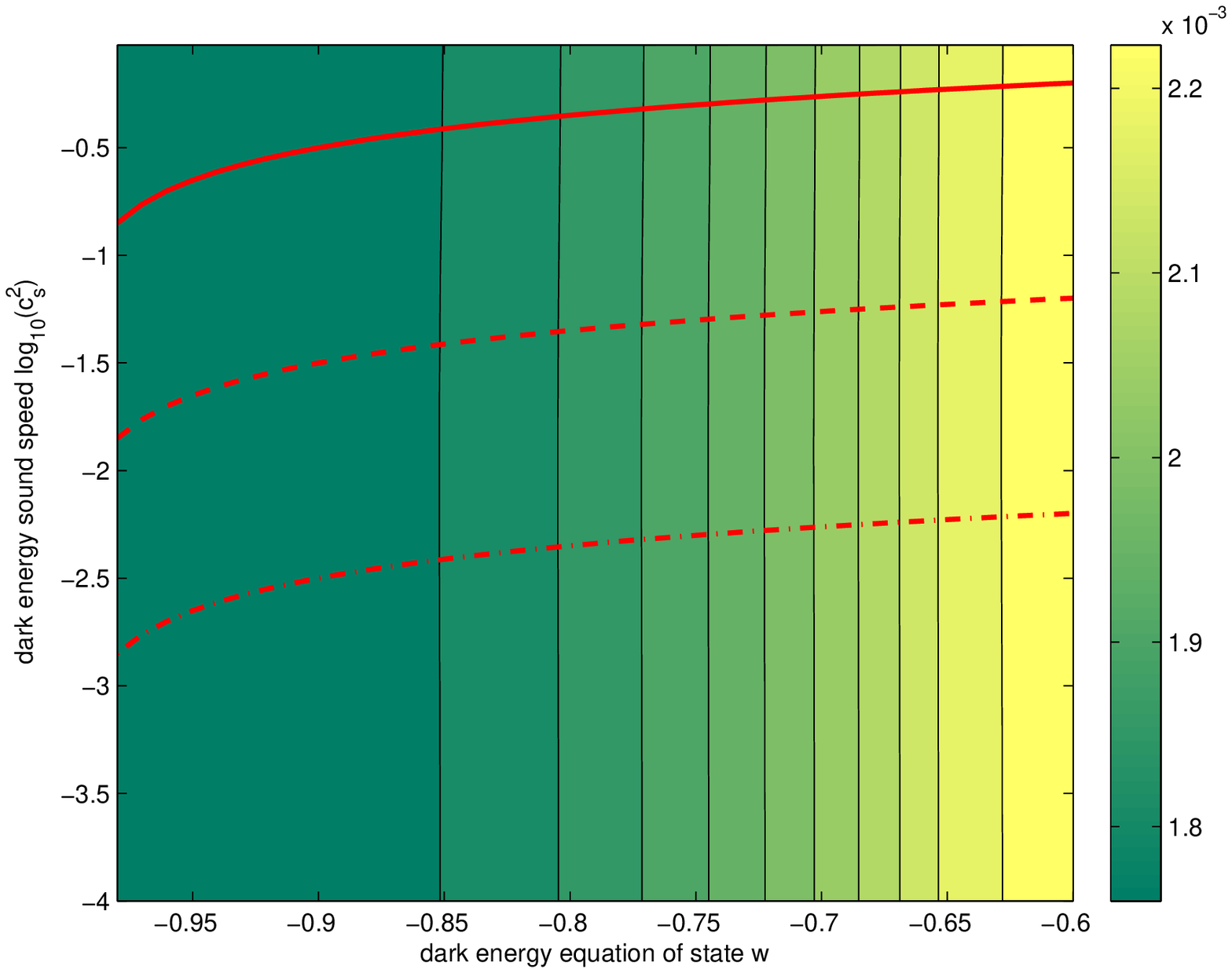}
	\end{center}
	\caption{Cram\'{e}r--Rao bounds on $\Delta \log_{10} c_s^2$ (upper figure) and $\Delta w
	/ |w|$ (lower figure) for varying fiducial values $w$ and
	$\log_{10} c_s^2$. The three red lines mark (from top to bottom)
	the scales $\lambda_\mathrm{eff} = 10^{-n}$, $n=1,2,3$ times the Hubble radius
	${\mathcal H}^{-1}$ according to Eq.~(\ref{eq:eff_scale_numerical}) in the
	$w$--$c_s$ plane.}
	\label{fig:w_cs2_plane}
\end{figure}

The Fisher
matrix is then a $2 \times 2$ matrix, and the uncertainties
are estimated as explained in Sec.~\ref{sec:parameter}. We combine
CMB and 3d weak lensing constraints. In our numerical calculation, the
multipoles $\ell$ run from $\ell_\mathrm{min} = 2$ to
$\ell_\mathrm{max} = 50$, the mode $k$ from $k_\mathrm{min} =
10^{-3}$~Mpc$^{-1}$ to $k_\mathrm{max} = 10^{-1}$~Mpc$^{-1}$ in $N_k =
200$ equidistant steps. The included redshift range is $z_\mathrm{min}
= 10^{-4}$ to $z_\mathrm{max} = 10$ in $N_z = 1000$ steps. For the CMB
Fisher matrix, we include, as in all subsequent calculations,
multipoles from $\ell = 2$ to $\ell = 2250$. The
uncertainties are shown in Fig.~\ref{fig:w_cs2_plane}. These results should be
taken as a first approximation due to the limitations of the Fisher formalism
when applied to weakly constrained parameters, \cfs Sec.~\ref{sec:parameter}. 

The lower figure, which shows the relative error $\Delta w/|w|$ on the
dark energy equation of state $w$, is easily interpreted. The
constraints on the equation of state parameter $w$ are largely
independent of the assumed sound speed $c_s^2$. So, at least, the
uncertainty in the sound speed $c_s^2$ does not worsen the accuracy
with which $w$ can be known, nor will a wrong assumption on $c_s^2$
introduce a significant bias on the estimate of $w$.

In the upper figure, we see that, conversely, the sound speed constraints
heavily depend on the fiducial values, as explained above.  For $w \gtrsim
-0.95$ and sufficiently small $c_s^2$ the estimated error $\Delta
\log_{10}c_s^2$ is smaller than one. We may thus hope that the combination of 3d
weak cosmic shear and the CMB will determine the order of magnitude of $c_s^2$.

For subsequent calculations, we choose the exemplary fiducial value 
$c_s^2 = 10^{-2}$. For $w$, the most natural choice, $w = -1$, is not adequate. If we
still chose $w$ close to $-1$, \eg, $w = -0.99$ or $w = -0.9$, all
results would strongly depend on the exact value chosen. Instead, we
decide to go further away from the observationally preferred value and
use $w = -0.8$ for illustration.

\begin{figure}
	\begin{center}
		\includegraphics[width=.5\textwidth]{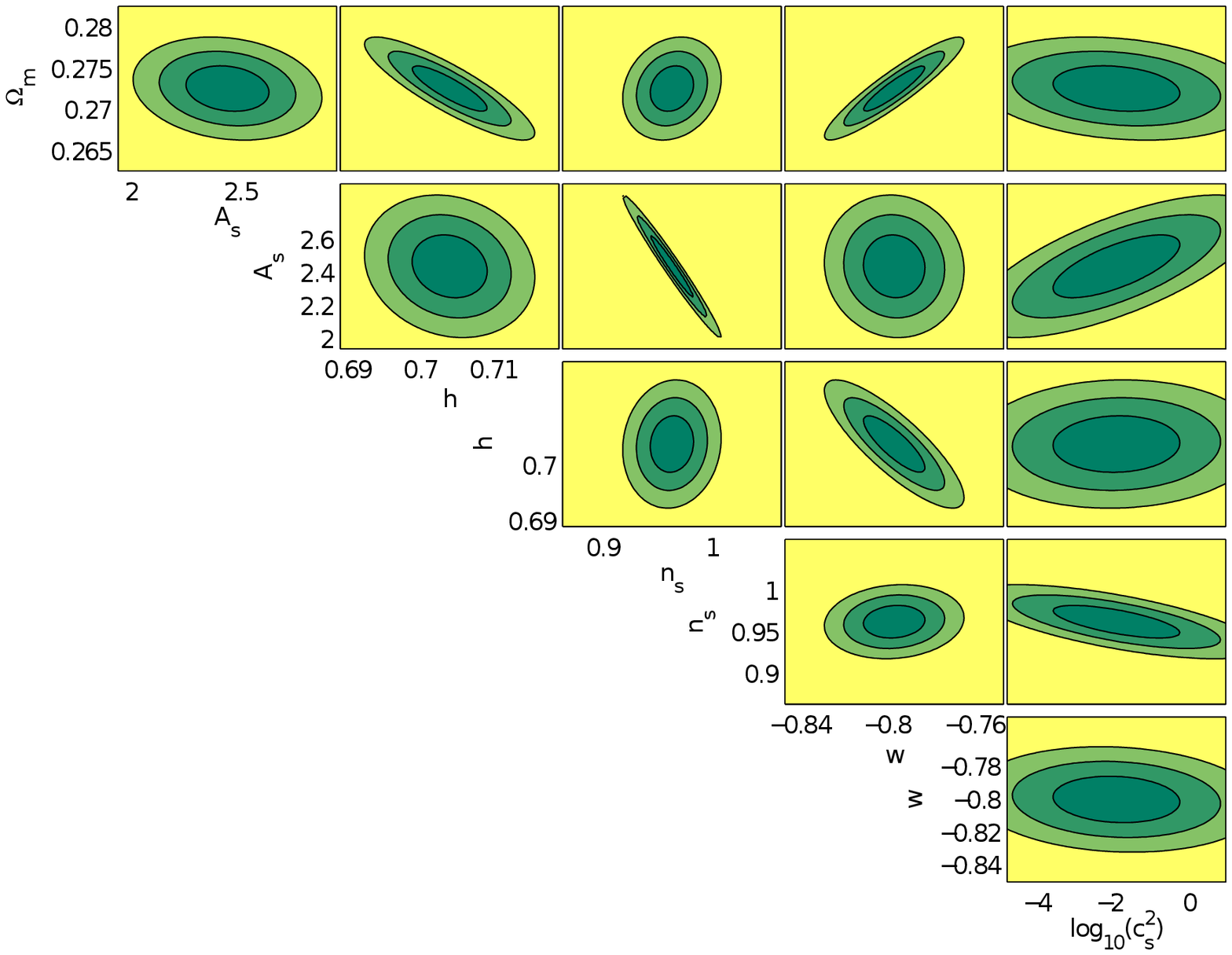}
		\includegraphics[width=.5\textwidth]{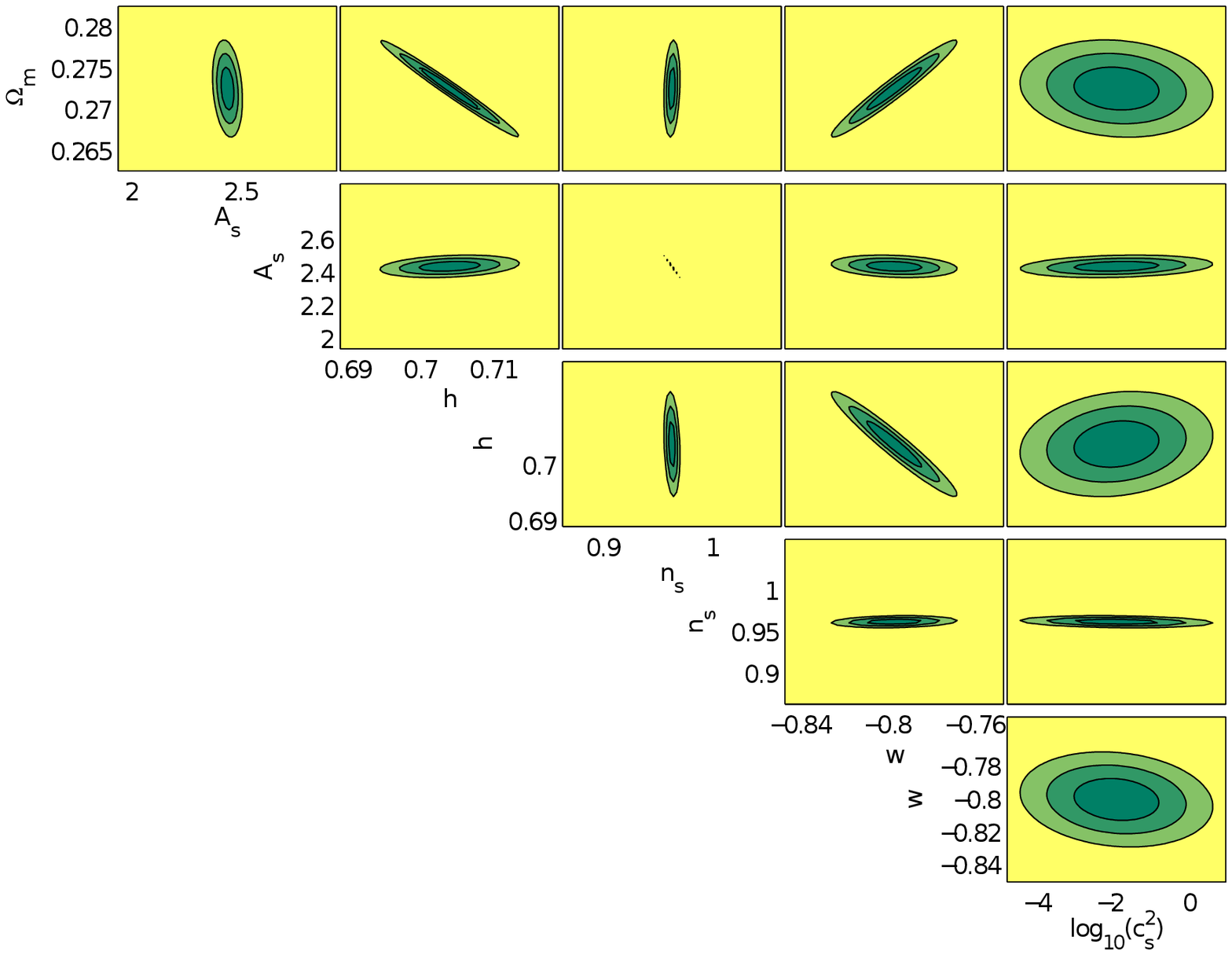}
	\end{center}
	\caption{Estimated confidence ellipses ($1\sigma$, $2\sigma$, and $3\sigma$) for
	3d weak lensing alone (upper figure) and combined with CMB prior
	(lower figure). The constraints on $A_s$ have been rescaled by a factor of $10^9$.}
	\label{fig:fisher}
\end{figure}

Adopting these choices for $c_s^2$ and $w$ as the fiducial values,
together with the $\Lambda$CDM WMAP7 recommended parameters
\citep{Komatsu10}, we now calculate the full Fisher matrices for our
six cosmological parameters: fractional matter density $\Omega_m$,
scalar initial perturbation amplitude $A_s$, Hubble parameter $h$,
scalar spectral index $n_s$, equation of state $w$, and sound speed
$\log_{10} c_s^2$. We choose higher numerical precision, $\ell_\mathrm{max} =
300$ and $N_k = 500$, and avoid nonlinear scales. The resulting
confidence regions for 3d weak cosmic shear alone and for the combined
constraints with the CMB are seen in Fig.~\ref{fig:fisher}.

3d weak cosmic shear obviously provides interesting constraints on all
the six cosmological parameters included in our analysis. 
Some constraints considerably improve when the CMB Fisher matrix is
added. This is not true for the sound speed.  
We emphasize, of course, that our choice of
fiducial parameters of $w$ and $c_s^2$ is only illustrative. In more
realistic cases $w \approx -1$, the constraints will be much weaker,
\cfs Fig.~\ref{fig:w_cs2_plane}.

\begin{figure}
	\begin{center}
		\includegraphics[width=.45\textwidth]{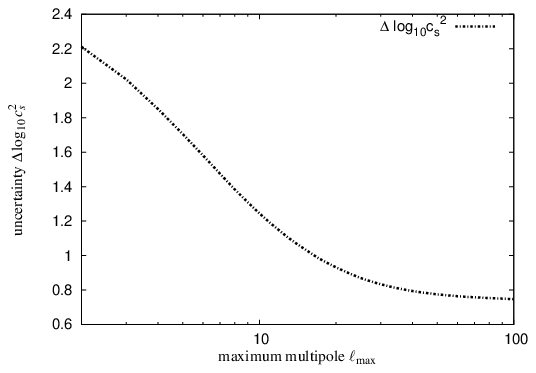}
		\includegraphics[width=.45\textwidth]{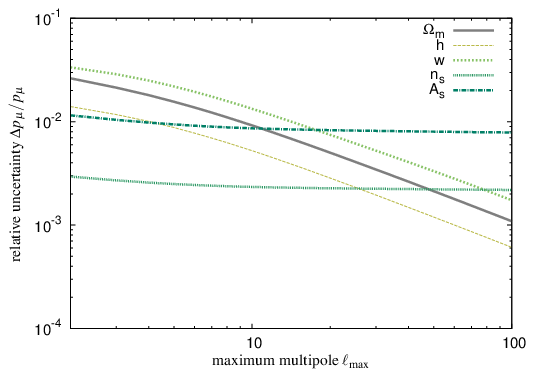}
	\end{center}
	\caption{Marginalised uncertainties $\Delta \log_{10} c_s^2$
	on the dark energy sound speed (top panel) and $\Delta
	p_\mu/p_\mu$ of the other cosmological parameters $p_\mu$ (bottom
	panel), depending on the maximum multipole $\ell_\mathrm{max}$
	from 3d weak lensing with a CMB prior. Fiducial parameters: $w =
	-0.8$, $c_s^2 = 10^{-2}$. The CMB prior generally includes all
	multipoles from $2$ to $2250$.}
	\label{fig:rel_l_errors}
\end{figure}

The Fisher matrix $F_{\mu\nu}$ for 3d weak lensing is obtained from a
summation of all multipoles $\ell$, \cfs
Eq.~(\ref{eq:fishercovariance}). It is instructive to examine which
multipoles most contribute to the parameter constraints. We therefore
plot the uncertainties of all the parameters as functions of
the maximum multipole $\ell_\mathrm{max}$ in Eq.~(\ref{eq:fishercovariance}),
see Fig.~\ref{fig:rel_l_errors}.

Let us first consider the parameters other than $c_s^2$. These show
two distinct behaviours. The constraints on the parameters $w$,
$\Omega_m$, and $h$ are strongly improved by going to larger
multipoles. The two parameters $A_s$ and $n_s$ characterising the
primordial scalar perturbation spectrum are already tightly
constrained for low multipoles. This is linked to the different
sensitivities of the two independent observations, 3d weak lensing and
the CMB, on these parameters.  Looking again at the error ellipses in
Fig.~\ref{fig:fisher}, we see that the constraints of 3d weak lensing
alone on $w$, $\Omega_m$, and $h$ are not much weaker than the
combined ones. Here, 3d weak lensing can establish strong constraints
with increasing $\ell_\mathrm{max}$. On the other hand, the CMB is
more sensitive to $A_s$ and $n_s$, whereby 3d weak lensing, regardless
of $\ell_\mathrm{max}$, cannot contribute very much to the
constraints.

\begin{figure*}
	\begin{center}
		\includegraphics[width=.3\textwidth]{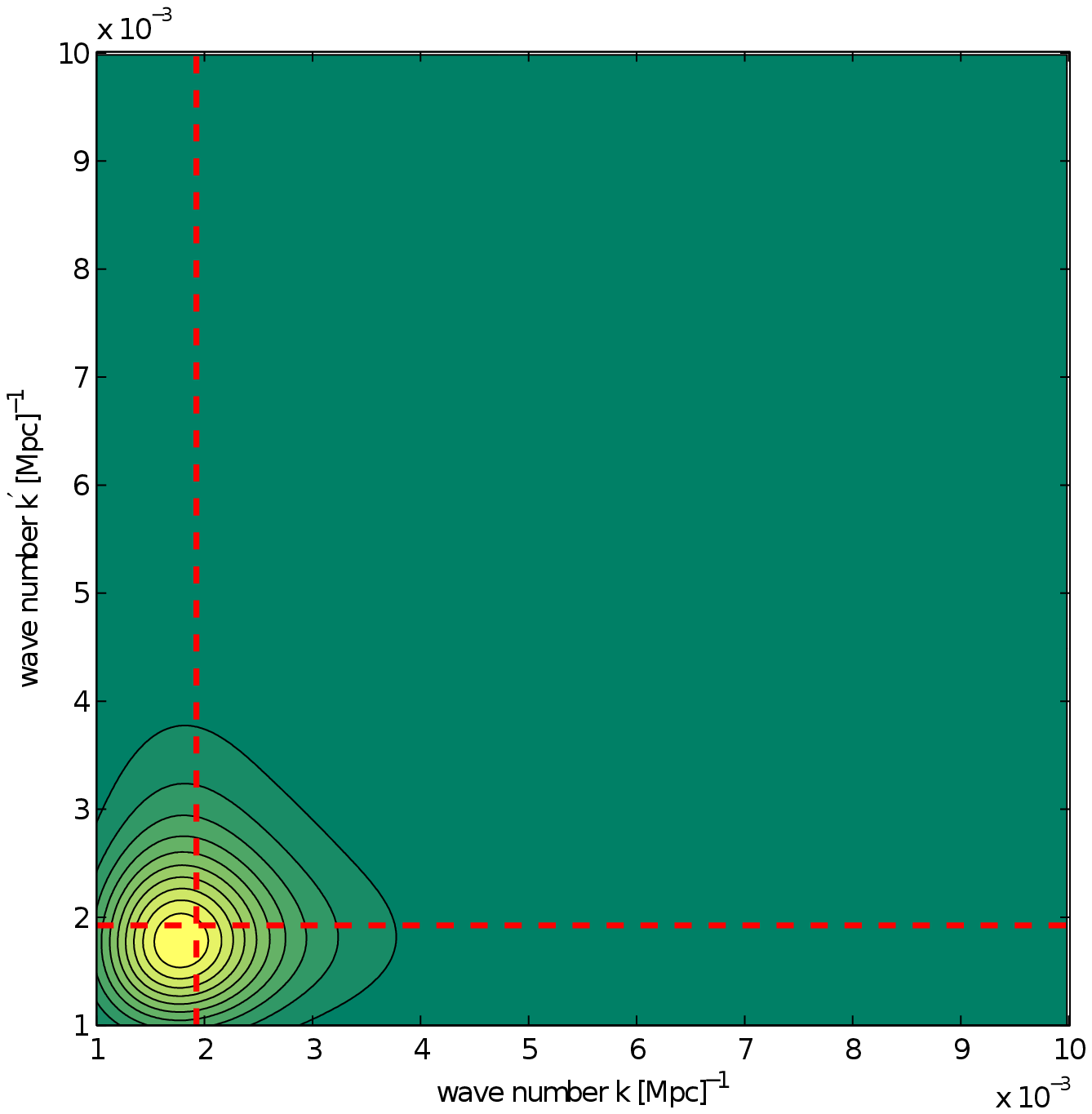}
		\includegraphics[width=.3\textwidth]{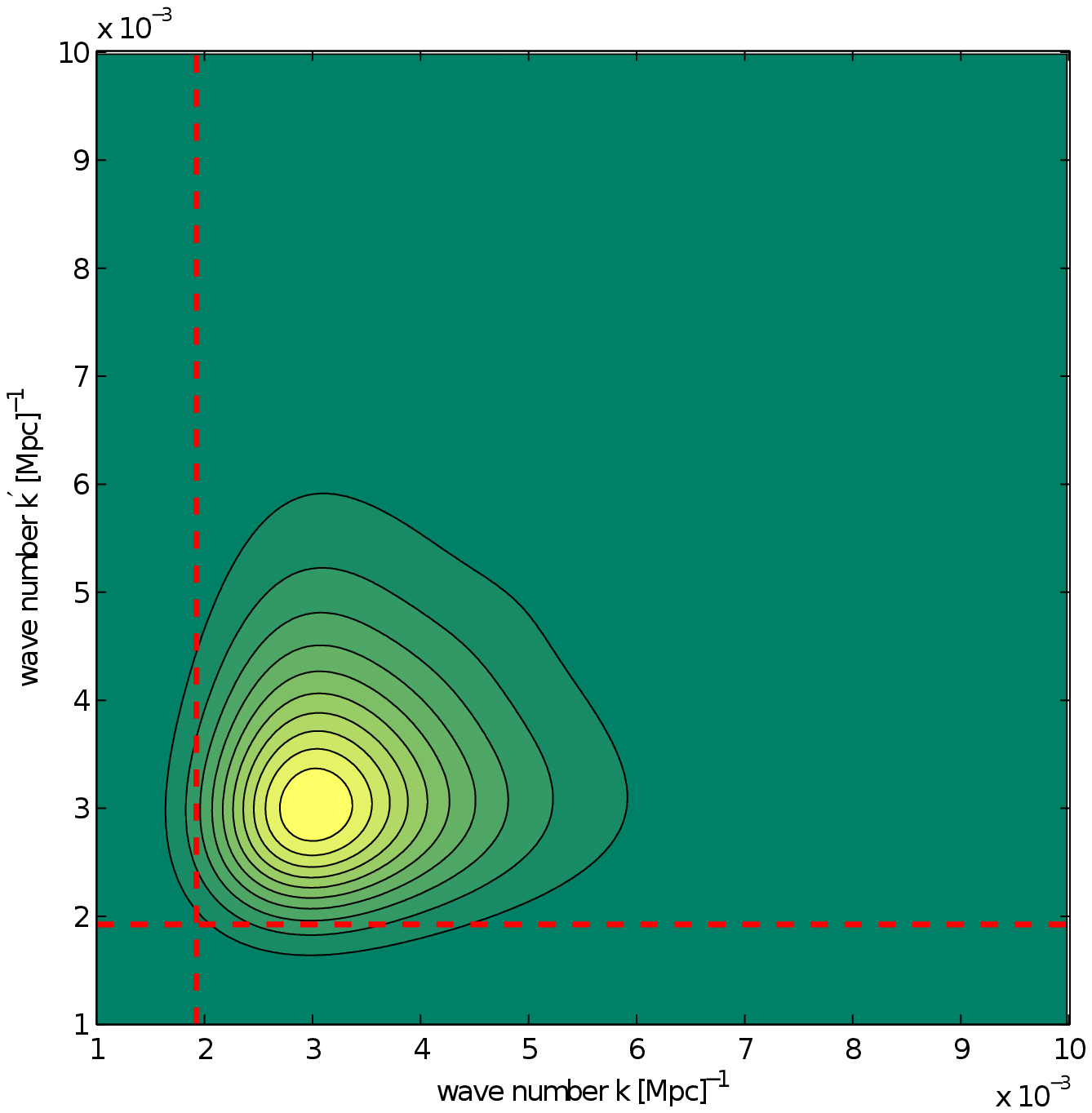}
		\includegraphics[width=.3\textwidth]{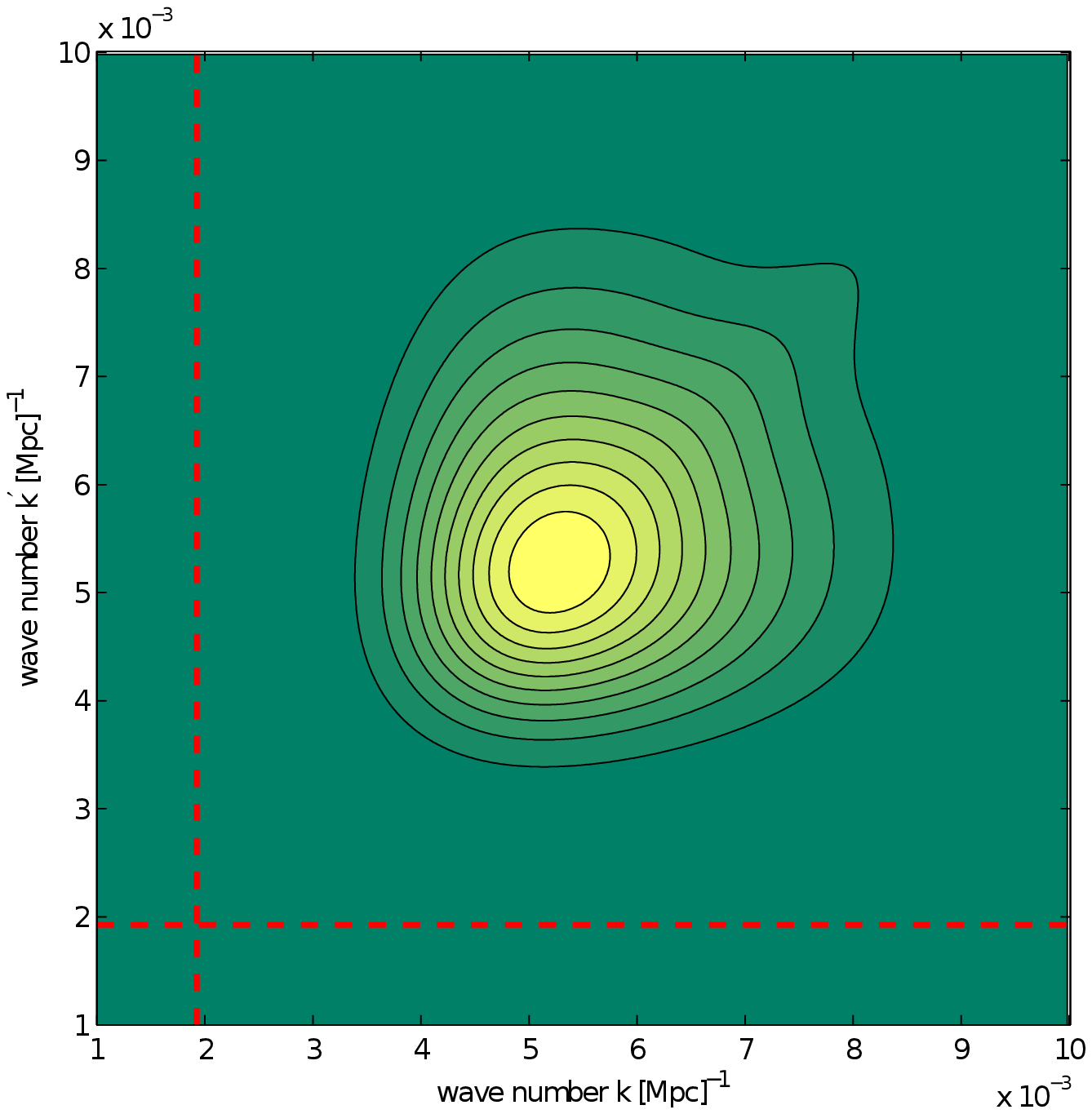}
	\end{center}
	\caption{Qualitative results for the covariance $C_\ell(k,k')$ for
	$\ell = 5,10,20$ and $w=-0.8$. Brighter regions mark larger
	values. The red dashed lines mark the scales $k_\mathrm{eff} = 1/
	\lambda_\mathrm{eff}$ in the case $c_s^2 = 1$.}
	\label{fig:covariances}
\end{figure*}

The case of the dark energy sound speed $c_s^2$ is different. The fact
that the uncertainty does not decrease significantly with
increasing $\ell_\mathrm{max} \gtrsim 20$ is mainly the consequence of clustering
dark energy being a large--scale phenomenon, \cfs
Sec.~\ref{sec:parameterised} and Fig.~\ref{fig:relative_spectra}.
Plotting the covariances $C_\ell(k,k')$ of the estimator, see
Sec.~\ref{sec:estimator}, for increasing multipoles $\ell$, we see how
the maximal sensitivity moves to smaller scales,
Fig.~\ref{fig:covariances}.
In fact, for low multipoles $\ell$, 3d weak shear probes the scales of interest
where dark energy clustering mainly occurs. The maxima seen in
Fig.~\ref{fig:covariances} are related to the fact that the galaxy distribution
$n(\chi)$, Eq.~(\ref{eq:nofz}), peaks at a comoving distance $\chi_*$
characterising the survey. Approximating Bessel functions by Dirac deltas, \cfs
Eq.~(\ref{eq:besselapprox}), this distance roughly corresponds to the scale $k \approx
\ell/\chi_*$. This explains the shift of the maximum for varying $\ell$
observed in Fig.~\ref{fig:covariances}.

Another way to study the $\ell$ dependence for the parameter
constraints is to look at the direct contribution of a multipole
$\ell$ to the diagonal elements $F_{\mu\mu}$ of the Fisher
matrix, Eq.~(\ref{eq:fishercovariance}). These quantities can be
interpreted as a (squared) sensitivity $s^\ell_\mu$ per $\ell$ mode,
\begin{equation}
	s^\ell_\mu \equiv 
	\mathrm{tr} \left( \mathbfss C_\ell^{-1} \partial_\mu \mathbfss C_\ell
	\right)^{2}
	\label{eq:sensitivity}
\end{equation}
Another way of interpreting Eq.~(\ref{eq:sensitivity}) is that
$s^\ell_\mu$ describes the derivative of the measurement with respect to a
cosmological parameter normalised by the noise of the measurement,
such that it assumes large values for strong dependences of the signal
on the cosmological model and small noise contributions. At the same
time, $s^\ell_\mu$ is the contribution to the Fisher matrix entry for
the parameter $p_\mu$ from each mode $\kappa_{\ell m}$ of the
convergence field. The number of modes for each $\ell$ is given by $2\ell +1$.

\begin{figure}
	\begin{center}		
		\includegraphics[width=.45\textwidth]{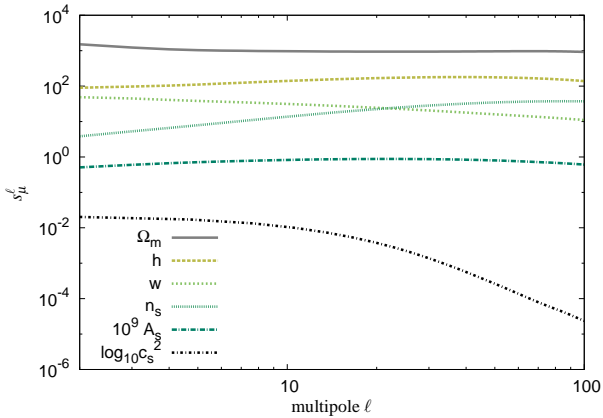}
	\end{center}
	\caption{Sensitivity $s^\ell_\mu$ as defined by
	Eq.~(\ref{eq:sensitivity}) for the cosmological parameters.
	Fiducial values: $w = -0.8$, $c_s^2 = 10^{-2}$.}
	\label{fig:sensitivity}
\end{figure}

We show the sensitivity $s^{\ell}_\mu$ for the cosmological parameters
in Fig.~\ref{fig:sensitivity}. Again, we observe that the sensitivity
of 3d weak cosmic shear on the dark energy sound speed $c_s^2$ mostly
comes from the first multipoles. This confirms the impression already
obtained from the covariances $C_\ell(k,k')$ shown in
Fig.~\ref{fig:covariances}, and emphasises the fact that the influence
of a nontrivial sound speed is a large--scale phenomenon. At the same
time, the plot explains the two orders of magnitude difference in
constraints on the dark energy sound speed compared to the other
cosmological parameters.

%
%

\section{Summary}
\label{sec:summary}

We have studied the potential of 3d weak cosmic shear to constrain a
possible clustering of dark energy with the data of next
generation surveys. We have parametrised the clustering dark energy
component by two parameters characteristic for a generic cosmological
fluid, its equation of state $w$ and its (rest--frame) sound speed
$c_s^2$. 

\begin{enumerate}
	\item{For the 3d weak cosmic shear analysis, we have developed
		adequate numerical tools allowing for an efficient calculation
		of the covariance and Fisher matrices. These tools were shown
		to be numerically very efficient, which ultimately allowed us
		to sweep through the dark energy parameter space, while
		retaining sufficient numerical accuracy. }
	\item{The capability of future observations to constrain $c_s^2$,
		and thereby the clustering of dark energy, strongly depends on
		the dark energy equation of state $w$. If $w$ is close to
		$-1$, dark energy perturbations are mainly present at very
		large scales possibly outside the Hubble horizon.  The effects of
		clustering dark energy would hardly be observable at all if
		$c_s^2 \gg 1 + w$.  As, indeed, current observations prefer
		values of $w$ very close to $-1$ \citep{Komatsu10}, this seems
		to be the decisive caveat.}
	\item{Due to the sensitivity of sound speed constraints to the
		assumed exact value of $w$, we have estimated the uncertainties 
		$\Delta \log_{10}c_s^2$ and $\Delta w / |w|$ as functions of
		the fiducial parameters $(w, c_s^2)$, \cfs
		Fig.~\ref{fig:w_cs2_plane}. The numbers are combined
		constraints based on assumed properties of Euclid
		\citep{Heavens03} and the Planck satellite \citep{Perotto06,
		Hollenstein09, Knox95}.}
	\item{For the considered range of fiducial parameters ($-0.99
		\lesssim w \lesssim -0.6$, $10^{-4}\lesssim c_s^2 \lesssim
		1$), the estimated constraints on the sound speed $\Delta
		\log_{10} c_s^2$ vary between the extreme cases of $0.1$ and $3$. 
		If $w\gtrsim -0.95$, the combination of Euclid and Planck is
		promising to constrain the order of magnitude of $c_s^2$ provided that
		the true sound speed is small enough.       
		This would be considerable progress compared not only to constraints possible with current
		observational data \citep{dePutter10, Li10} but also,
		for most of the parameter space, to the constraints expected from weak lensing
		tomography and galaxy surveys alone \citep{Sapone10}.
		Constraining $c_s^2$ within one or two orders of magnitude
		could also be possible with Planck and next--generation galaxy
		surveys \citep{Takada06, Ballesteros10} or for neutral
		hydrogen surveys \citep{TorresRodriguez07, TorresRodriguez08}.
		Although not our focus here, the constraints of 3d weak cosmic
		shear together with the Planck satellite on the dark energy
		equation of state $w$ are worth mentioning; in fact, according
		to Fig.~\ref{fig:w_cs2_plane} and Fig.~\ref{fig:fisher}, $w$
		can be constrained below the percent level \citep[for other
		constraints from 3d weak cosmic shear, see][]{Heavens03,
		Heavens06}. Additionally, the true value of $w$ largely
		determines the accuracy on the sound speed $c_s$.}
\end{enumerate}

There are two very different conclusions one could draw from these
constraints on $c_s^2$. First, we may regard $c_s^2 = 1$ and $w
\approx -1$ as the natural values as they refer to unclustered dark
energy such as a cosmological constant. Then, small deviations from
$c_s^2 = 1$ are interesting; but these seem hardly observable in next
generation experiments. Second, however, one may argue that $c_s^2$
is a completely unknown parameter with a natural range from $0$ to
$1$; then, Euclid and 3d weak lensing could single out an order of
magnitude in which $c_s^2$ lies. This could be a decisive step for
discriminating between different dark energy models.

\section*{Acknowledgements} MW and YA acknowledge support from the DFG
Transregional Collaborative Research Centre TRR~33, and BMS's work was
supported by the German Research Foundation (DFG) within the framework
of the excellence initiative through the Heidelberg Graduate School of
Fundamental Physics. We would like to thank Matthias Bartelmann,
Alessandra Grassi, and Angelos Kalovidouris for their suggestions.

\bibliography{aamnem,sound}
\bibliographystyle{mn2e}

\bsp

\label{lastpage}

\end{document}